\documentclass[aps,prl,twocolumn,showpacs, floatfix]{revtex4}
\usepackage{amsmath,amssymb}
\usepackage[dvips]{graphicx}

\DeclareMathOperator{\Sp}{Tr}

\begin{document}

\title{Truncated-Determinant Diagrammatic Monte Carlo
for Fermions with Contact Interaction}
\author{Evgueni Bourovski}
\author{Nikolay Prokof'ev}
\author{Boris Svistunov}
\affiliation{Physics Department, University of Massachusetts,
Amherst, USA, 01003} \affiliation{Russian Research Center
``Kurchatov Institute'', 123182 Moscow }
\begin{abstract}
For some models of interacting fermions the known solution to the
notorious sign-problem in Monte Carlo (MC) simulations is to work
with macroscopic fermionic determinants; the price, however, is
a macroscopic scaling of the numerical effort spent on elementary
local updates. We find that the {\it ratio} of two macroscopic determinants
can be found with any desired accuracy by considering truncated (local in
space and time) matices. In this respect, MC for interacting fermionic
systems becomes similar to that for the sign-problem-free bosonic
systems with system-size independent update cost. We demonstrate
the utility of the truncated-determinant method by simulating
the attractive Hubbard model within the MC scheme based on partially summed
Feynman diagrams. We conjecture that similar approach may be
useful in other implementations of the sign-free determinant
schemes.
\end{abstract}
\maketitle


Monte Carlo (MC) methods are a unique tool for studying large
interacting systems. The most severe limitation on their
applicability is imposed by the so-called sign-problem (SP) when
relevant contributions to statistics alternate in sign and almost
exactly compensate each other in the final answer \cite{book}.
Frustrating interactions and anticommutation relations for fermion
operators are typically at the origin of the sign-problem. In this
Letter, we address the case when quantum statistics is
non-positive only because of the fermion exchange cycles.

One solution to the fermion SP is offered by the determinant Monte
Carlo (DetMC) (see, e.g. \cite{Loh92}). The idea is that all
contributions to the many-body statistics obtained by exchanging
fermion places (in a certain representation) can be written as a
product of two determinants---for spin-up and spin-down
species---and in cases when the two real determinants coincide
the result is positive definite. In Metropolis-type algorithms
\cite{Metr53}, MC updates are accepted with
probabilities proportional to the ratio of final and initial
configuration weights; in DetMC the corresponding acceptance
ratio, $R$, is based on the ratio of large determinants.
Unfortunately, calculating determinants ratio for macroscopically
large matrices is very expensive numerically: even with
tricks involving the Hubbard-Stratonovich tranformation the
algorithm proposed by Blankenbecler, Scalapino
and Sugar \cite{ScaSu@PRL81,BSS81} still requires $L^{2d}$
operations per update for a $d$-dimensional system with $L$
lattice points per dimension. The same scaling is true for the
continuous-time scheme \cite{Romb99}. In contrast, for bosonic
systems with local interactions the number of operations per
update is small and system-size independent, i.e. they can be
simulated $L^{2d}$ times faster!

Since the bottleneck of DetMC is the calculation of $R$, one may
question the paradigm of calculating it ``exactly'': In any case,
computer operations always involve systematic round-off errors.
The other example is provided by (pseudo)random number
generators---they are {\it always} imperfect and result in
systematic errors equivalent to small errors in $R$ which,
however, remain practically undetectable (for good generators) in
final results. The heuristic explanation of why small errors in
$R$ do not ruin the simulation is as follows. The Metropolis
algorithm is a scheme with strong relaxation towards equilibrium
distribution, and local configuration updates may be viewed as the
result of dissipative coupling to the thermal bath (this picture
is often used to model dissipative kinetics \cite{book}).
Uncontrolled errors in $R$ may then be regarded as a small
stochastic noise in the relaxational dynamics. As such, it only
slightly modifies the equilibrium state and its properties. This
is a standard argument in the linear response theory.

It seems natural then to suggest that if the goal is to simulate
the result with $n$-digit accuracy, there is no need to calculate
the acceptance ratio with accuracy much higher than $n$ digits.
Often, we simply ignore this issue because getting $R$ with
machine precision does not cost any extra CPU time. In determinant
methods, however, there is a potential of huge efficiency gains if
approximate values of $R$ can be calculated much faster. We
demonstrate the feasibility of this approach by showing that the
ratio of two macroscopic determinants can be found with high
accuracy by considering truncated matrices dealing only with the
local (in space-imaginary time) structure of the configuration
space. The computational cost of updates in the corresponding
``truncated-determinant'' scheme is system-size independent---an
efficiency increase  $\propto L^{2d}$ for large $L$.

In what follows, we discuss the solution of the Hubbard model for
fermions within a simple diagrammatic MC scheme based on Feynman
diagrams partially summed over fermion propagator permutations
\cite{Romb99,Rub}. The resulting diagram weight is the square of
the determinant composed of finite-temperature fermion
propagators. We explain how the determinant ratio for local
updates may be calculated using truncated matrices, and
demonstrate the feasibility of the proposed approach. We also show
that going to larger system sizes has little effect on the scheme
performance. Finally, we conjecture that large efficiency gains
are expected in other sign-problem free DetMC schemes, e.g. in
lattice QCD simulations with quark fields \cite{QCD}. It is also
worth noting that in the diagrammatic DetMC scheme the update cost
does not depend directly on the lattice period, which is a big
advantage for simulations of dilute systems. By ``dilute" we mean
dilute with respect to the lattice, but not necessarily with
respect to the interaction; the latter can be effectively large
due to a resonance on a (quasi)-bound state. Correspondingly, the
new scheme is very promising for the study of ultra cold fermionic
systems in the regime of strong Feshbach resonant interaction,
including the crossover from the Bardeen-Cooper-Schrieffer pairing
to the Bose-Einstein condensation of molecules \footnote{Barbara
Goss Levi, Physics Today \textbf{56}, 18 (2003), and references
therein; for recent results see D. S. Petrov, C. Salomon and G. V.
Shlyapnikov, cond-mat/0309010; M. Greiner, C. A. Regal, and
Deborah S. Jin, Phys. Rev. Lett. \textbf{92}, 040403 (2004); S.
Jochim \textit{et al.}, Science, \textbf{32}, 2101 (2003); M. W.
Zwierlein \textit{et al.}, Phys. Rev. Lett. \textbf{91}, 250101
(2003).}.


\textit{Model and method.} We consider interacting
lattice fermions with the Hubbard Hamiltonian $H = H_0 + H_1$:
\begin{equation}
H_0 = \sum_{\mathbf{k}\sigma} \, (\epsilon_\mathbf{k} -
\mu)c^{\dagger}_{\mathbf{k}\sigma} c_{\mathbf{k}\sigma}, \; \;\;\;
H_1 = U\sum_{\bf x}n_{{\bf x} \uparrow}n_{{\bf x} \downarrow}, %
\label{ham_1}
\end{equation}
where $c^{\dagger}_{{\bf x} \sigma}$ is the fermion creation
operator, $n_{{\bf x}\sigma} = c^{\dagger}_{{\bf x} \sigma}c_{{\bf
x}\sigma}$, $\sigma =\, \uparrow, \downarrow$ is the spin index,
${\bf x}$ runs over the $L^d$ points of the simple cubic lattice,
$\mathbf{k}$ runs over the corresponding Brillouin zone,
$\epsilon_\mathbf{k} =-2t \sum_{\alpha = 1}^d \cos k_{\alpha} a $
is the tight-binding dispersion law, and $\mu$ is the chemical
potential. For definiteness and numerical tests, we confine
ourselves to the $d=2$ spacial lattice with periodic boundary
conditions. We use the hopping amplitude, $t$, and lattice
constant $a$ as units of energy and distance, respectively.

Following Refs.~\cite{Romb99,Rub} we start with writing
the statistical operator in the interaction representation,
\begin{equation}
e^{-\beta H } = e^{-\beta H_0 } \mathcal{T}_{\tau }\exp
\{-\int_0^\beta H_1 (\tau ) d\tau  \}
\end{equation}
where $H_1 (\tau ) = e^{H_0 \tau } H_1 e^{-H_0 \tau }$ and
$\mathcal{T}_{\tau }$ is the time ordering operator, and expanding
it in powers of $H_1$:
\begin{eqnarray}
Z &=& \sum_{p=0}^{\infty} (-U)^p  \sum_{{\bf x}_1 \dots  {\bf x}_p
 } \int_{\tau_1 < \tau_2 <\dots <\tau_p< \beta}
\left( \prod_{i=1}^p d\tau_i \!\right) ~~~~~~~~~~ \nonumber \\
&\times & \Sp \left[ e^{-\beta H_0 }
 \prod_{i=1}^p c^{\dagger}_{\uparrow}(\mathbf{x}_i\tau_i)
c_{\uparrow}(\mathbf{x}_i\tau_i)
c^{\dagger}_{\downarrow}(\mathbf{x}_i\tau_i)
c_{\downarrow}(\mathbf{x}_i\tau_i) \right] .
\label{Z}
\end{eqnarray}
This expansion for the partition function generates
standard Feynman diagrams \cite{Fetter-Walecka}.
Graphically, each term is a set of
four-point vertices with two incoming (spin-up and spin-down), and
two outgoing (spin-up and spin-down) lines which connect vertices.
Each line is associated with the imaginary time fermion
propagator, $G_{\sigma}(\mathbf{x}_i-\mathbf{x}_j,\tau_i-\tau_j;
\mu,\beta) = - \Sp \left[ \mathcal{T}_\tau e^{-\beta H_0}
c_{\sigma}(\mathbf{x}_i\tau_i)
               c^{\dagger}_{\sigma}(\mathbf{x}_j\tau_j) \right]$.

A straightforward MC sampling of diagrammatic series would be
impossible because of the sign-problem. However, if for a given
configuration of $p$ vertices,
$\mathcal{S}_p = \{ ( \mathbf{x}_j,\tau_j ) ,\; j=1,\ldots,p \} $,
one sums over all $(p!)^2$ ways of connecting them by propagators,
then the result can be written as a {\it product of two
determinants}, one for spin up, and another for spin down (see
e.g. \cite{Loh92}). The differential weight of the vertex
configuration (or vertex diagram) is then
\begin{equation}
d\mathcal{P}(\mathcal{S}_p)= (-U)^p   \, \det\mathbf{A}^{\uparrow}
(\mathcal{S}_p) \det\mathbf{A}^{\downarrow}(\mathcal{S}_p)\,
\prod_{i=1}^p d\tau_i \;, \label{P}
\end{equation}
where
$\mathbf{A}^{\sigma}(\mathcal{S}_p)$ are $p\times p$ matrices:
$A^{\sigma}_{ij}= G_{\sigma}(\mathbf{x}_i-\mathbf{x}_j,\tau_i-\tau_j)$.
For equal number of up- and down-particles, $ \det\mathbf{A}^{\uparrow}
\det\mathbf{A}^{\downarrow} = [\det\mathbf{A} ]^2$, and negative
$U$ the vertex diagram weight is always positive. [At half
filling, $n_{\uparrow}+n_{\downarrow}=1$, the sign of $U$ changes
when hole representation is used for one of the spin components,
so this scheme may be also used for the repulsive Hubbard model.]

MC sampling in the vertex configuration space $(p,\mathcal{S}_p)$
can be performed by standard MC rules (see, e.g.
Refs.~\cite{DMC,Rub}) using just one pair of complementary updates
${\cal D}$ and ${\cal C}$: in ${\cal D}$ one selects at random one
of the vertices and suggests to delete it from the configuration;
in ${\cal C}$  an additional vertex is suggested to be inserted at
some point randomly selected in the space-time box $\beta \times
L^d$. These updates decrease/increase the rank of $\mathbf{A}$ by
one. The acceptance ratio for the ${\cal D}$/${\cal C}$ pair of
updates is then based on the ratio of two determinants
\begin{equation}
R_{p} =  { \det \mathbf{A}(\mathcal{S}_{p+1}) /
           \det \mathbf{A}(\mathcal{S}_{p}) } \;,
\label{upd_ratio}
\end{equation}
where $ \mathcal{S}_{p+1} = \{ \mathcal{S}_p ,(\mathbf{x}_{p+1},
\tau_{p+1}) \} $ (we omit the spin index for brevity).

%

The bottleneck of this simple scheme is in evaluating $R_{p}$ when
$p$ is macroscopically large.  The typical number of vertices is
determined by the number of particles, interaction strength, and
inverse temperature as $p \propto N \beta U $. The
truncated-determinant idea is to calculate $R_p$ much faster at
the expense of accuracy using the following conjecture originating
from physical, rather than mathematical, arguments. The vertex
configuration represents a sequence of virtual particle collisions
in the many-body system, and it is likely that {\it local} changes
in its structure depend only on the immediate neighborhood of the
updated region. [We note that the idea of employing the local
nature of the fermion-boson coupling has been used in
\cite{ScaSu@PRL81}, but it has not been extended to the fermionic
determinant.] Quantitatively, we define a norm, $\| \dots \|$, or
a distance, between vertices in space-time (several choices are
discussed below), and construct a truncated vertex configuration,
$\mathcal{S}_p^{(\ell)}$, such that all points in
$\mathcal{S}_p^{(\ell)}$ satisfy
\begin{equation}
\| (\mathbf{x}_j,\tau_j) - (\mathbf{x}_{p+1},\tau_{p+1}) \|
\leqslant \ell  \;. \label{local}
\end{equation}
Correspondingly, $ \mathcal{S}_{p+1}^{(\ell)} = \{
\mathcal{S}_p^{(\ell)} ,(\mathbf{x}_{p+1} \tau_{p+1}) \} $.
We may now use truncated configurations to calculate
the ratio (\ref{upd_ratio}) approximately as
\begin{equation}
R_p^{(\ell)} = \det \mathbf{A}( \mathcal{S}_{p+1}^{(\ell)} )
/ \det \mathbf{A}(\mathcal{S}_{p}^{(\ell)}).
\label{ratio_cut}
\end{equation}
Clearly,  when $\ell \to L$ we recover the exact ratio. Our
conjecture is then that $R_p^{(\ell)}$ quickly converges to $R_p$
and there exists a healing length in the $(\mathbf{x},\tau)$-space
characterizing this convergence. If this is the case, then $\ell $
may be considered as a microscopic (system-size independent)
parameter controlling the accuracy and efficiency of simulation.

The proper choice of the norm $\| \dots \|$ depends on system
parameters. In the strongly correlated case
the natural units of distance and time are provided
by the Fermi momentum, $k_F$, and Fermi energy $\epsilon_F$.
One possibility is then
\begin{equation}
\|(\mathbf{x},\tau) - (\mathbf{x}',\tau ' ) \| =
\sqrt{
 k_F^2 ( \mathbf{x} - \mathbf{x}') ^2 +
 \epsilon_F^2 (\tau -\tau') ^2
 } \;.
\label{norm_sph}
\end{equation}
Geometrically, this measure results  in a set of vertices
$\mathcal{S}_{p}^{(\ell)}$ inside the  space-time ellipsoid
centered at $(\mathbf{x}_\mathrm{p+1},\tau_\mathrm{p+1})$. For
dense systems $\|(\mathbf{x},\tau) - (\mathbf{x}',\tau') \| = \max
\{ |\mathbf{x}-\mathbf{x}'| , |\tau - \tau'| \}$ is equally
appropriate and our data for the largest system were obtained
using this measure. At temperatures comparable to $\epsilon_F$ one
may account for all vertices in the $\hat{\tau }$-direction and
simply write $ \|(\mathbf{x},\tau) - (\mathbf{x}',\tau')
\|_\mathrm{cyl} = k_F | \mathbf{x} - \mathbf{x}' | $. The
corresponding geometrical figure is a $\beta$-cylinder. Similarly,
in small systems one may consider truncating configurations only
in time direction.

\begin{figure}[htb]
\includegraphics[width = 0.99\columnwidth,keepaspectratio=true]{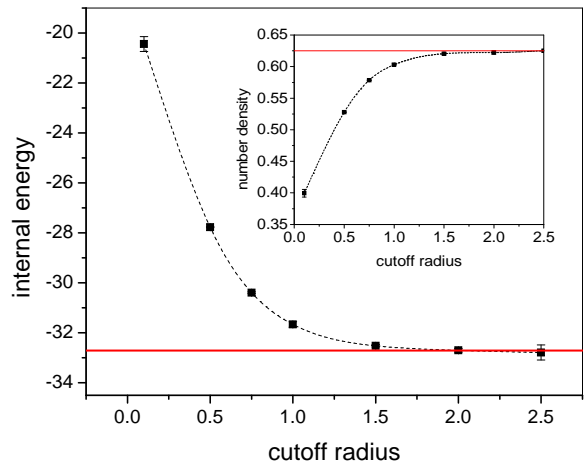}
\caption{Energy and density dependence on the imaginary
time cutoff for the Hubbard model for $L=4$.
Dots represent the MC data, and solid lines are the exact
diagonalization results for 10 particles \cite{Husslein97}.}
\label{fig:Hubb}
\end{figure}

\textit{Numerical results and discussion.}
Our tests of the truncated-determinant scheme were done for
the attractive Hubbard model with $U=-4$, $\mu=-2$, $\beta=10$,
and periodic boundary conditions.
First, we simulated a small $L^2= 4^2$ cluster
for which the ground state energy (of 10 particles) is known from
exact diagonalization studies \cite{Husslein97}. Since the spatial
dimension is so small we truncate vertex configurations only in the
imaginary time direction.
In Fig.~\ref{fig:Hubb} we show how the result for energy converges
to the exact value. We stress, that at all stages of the MC simulation
we never even write the full configuration determinant, which rank
is about $2.5$ times larger than typical values of $p$ for the
cutoff radius $2$.

\begin{figure}[htb]
\includegraphics[width = 0.99\columnwidth,keepaspectratio=true]{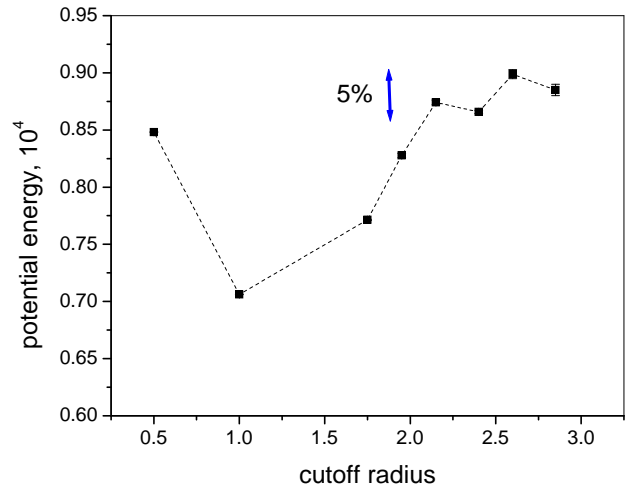}
\caption{Potential energy dependence on the truncation
radius for the Hubbard model for $L=100$.}
\label{fig:100}
\end{figure}

\begin{figure}[htb]
\includegraphics[width = 0.99\columnwidth,keepaspectratio=true]{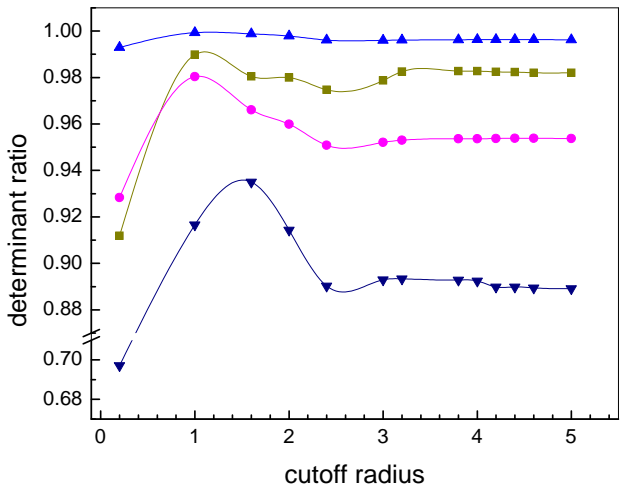}
%
\caption{Determinant ratios $R_{p}^{(\ell)}$ as functions of
$\ell$  for randomly selected MC configurations for the $L=100$
system.} \label{fig:ratios}
\end{figure}

In Fig.~\ref{fig:100} we present our data for the $L^2=100^2$
system---now, using determinant truncation both in space and in
time directions, Eq.~(\ref{norm_sph}). Remarkably, the convergence
is achieved around the same value of the truncation radius, which
proves that the computational cost per update is not subject to
macroscopic scaling.  It is instructive to see how data
convergence for energy correlates with the typical errors
introduced by the approximate calculation of $R_p$. In
Fig.~\ref{fig:ratios} we show examples of $R_p$ dependence on the
truncation radius for a number of randomly selected MC
configurations. Clearly, quite large fluctuations in $R_p$ are
statistically ``averaged out'' in the final result for energy. We
are not aware of any other method capable of simulating fermionic
systems of comparable size.

Apart from the Hubbard model tests, we have also verified that the
use of truncated-determinants for randomly seeded vertex
configurations works as nicely to speed up the calculation of
$R_p$.

The benchmark DetMC method by Blankenbecler, Scalapino and Sugar
(BSS) \cite{ScaSu@PRL81,BSS81} is based on the Trotter-Suzuki
imaginary time slicing, and the Hubbard-Stratonovich
transformation \cite{Hirsch83}. The rank of the matrix used in the
calculation of the acceptance ratio equals the number of lattice
sites $L^{d}$, and $\sim L^{2d}$ operations are required to find
its determinant. The necessity of handling large matrices,
although made possible by this method, requires both elaborate
finite-size scaling analysis of MC data \cite{Sewer02}, and
special efforts for the calculation stabilization at low
temperatures \cite{Loh92}. The contour-distortion stabilization
techniques (see, e.g. \cite{Rom97,Baeurle02}) help to alleviate
the sign problem, but still suffer from the severe scaling of the
computational cost per update. More recently, Rombouts, Heide, and
Jachowicz \cite{Romb99} improved the BSS scheme by formulating it
in the $\tau$-continuum. It is easy to directly compare this
scheme with ours because the starting point is exactly the
same---the expansion of the statistical operator in powers of $U$.
Rombouts {\it et al.} used the auxiliary Ising variables to
decompose four-point vertices into the sums of single-particle
exponentials, and arrived at the number of operations for
performing one update scaling as $L^{2d}$. At this point we notice
that while we work with the same vertex configuration structure,
in our scheme there is no extra summation over the auxiliary
variables, and the calculation of the acceptance ratio is
system-size independent.

Recently, a substantial improvement in lattice QCD simulations has
been achieved by including quark-loop effects (see, e.g.
\cite{QCD}). After the quarks are integrated out, their effects
are described by the macroscopic positive definite determinant,
$\det \mathbf{A}(U)$, where $U$ is the configuration of gluon
matrices. The conjecture is that the ratio $\det
\mathbf{A}(U)/\det \mathbf{A}(U')$ for local updates of gluon
matrices, $U \to U'$, can be calculated by accounting only for the
immediate neighborhood of the updated lattice bond.

We doubt that the truncated-determinant schemes will help to speed
up simulations with the sign-problem. If the average configuration
sign is small, the answer is determined by small differences
between the sign-positive and sign-negative contributions, i.e.
each contribution has to be calculated to much higher accuracy
than would be sufficient in the positive definite case. As a
result, higher and higher precision is required for $R_p$, and
advantages of our approach quickly vanish.

\bibliography{detMC_2}

\begin{thebibliography}{15}
\expandafter\ifx\csname natexlab\endcsname\relax\def\natexlab#1{#1}\fi
\expandafter\ifx\csname bibnamefont\endcsname\relax
  \def\bibnamefont#1{#1}\fi
\expandafter\ifx\csname bibfnamefont\endcsname\relax
  \def\bibfnamefont#1{#1}\fi
\expandafter\ifx\csname citenamefont\endcsname\relax
  \def\citenamefont#1{#1}\fi
\expandafter\ifx\csname url\endcsname\relax
  \def\url#1{\texttt{#1}}\fi
\expandafter\ifx\csname urlprefix\endcsname\relax\def\urlprefix{URL }\fi
\providecommand{\bibinfo}[2]{#2}
\providecommand{\eprint}[2][]{\url{#2}}

\bibitem[{\citenamefont{Binder and Landau}(2000)}]{book}
\bibinfo{author}{\bibfnamefont{K.}~\bibnamefont{Binder}} \bibnamefont{and}
  \bibinfo{author}{\bibfnamefont{D.~P.} \bibnamefont{Landau}},
  \emph{\bibinfo{title}{A Guide to Monte Carlo Simulations in Statistical
  Physics}} (\bibinfo{publisher}{Cambridge University Press},
  \bibinfo{address}{Cambridge}, \bibinfo{year}{2000}).

\bibitem[{\citenamefont{Loh and Gubernatis}(1992)}]{Loh92}
\bibinfo{author}{\bibfnamefont{E.~Y.} \bibnamefont{Loh}} \bibnamefont{and}
  \bibinfo{author}{\bibfnamefont{J.~E.} \bibnamefont{Gubernatis}}, in
  \emph{\bibinfo{booktitle}{Electronic Phase transitions}}, edited by
  \bibinfo{editor}{\bibfnamefont{W.}~\bibnamefont{Hanke}} \bibnamefont{and}
  \bibinfo{editor}{\bibfnamefont{Y.~V.} \bibnamefont{Kopaev}}
  (\bibinfo{publisher}{Elsevier Science Publishers}, \bibinfo{address}{New
  York}, \bibinfo{year}{1992}).

\bibitem[{\citenamefont{Metropolis et~al.}(1953)\citenamefont{Metropolis,
  Rosenbluth, Rosenbluth, Teller, and Teller}}]{Metr53}
\bibinfo{author}{\bibfnamefont{N.}~\bibnamefont{Metropolis}},
  \bibinfo{author}{\bibfnamefont{A.}~\bibnamefont{Rosenbluth}},
  \bibinfo{author}{\bibfnamefont{M.}~\bibnamefont{Rosenbluth}},
  \bibinfo{author}{\bibfnamefont{A.~H.} \bibnamefont{Teller}},
  \bibnamefont{and} \bibinfo{author}{\bibfnamefont{E.}~\bibnamefont{Teller}},
  \bibinfo{journal}{J. Chem. Phys.} \textbf{\bibinfo{volume}{21}},
  \bibinfo{pages}{1087} (\bibinfo{year}{1953}).

\bibitem[{\citenamefont{Scalapino and Sugar}(1981)}]{ScaSu@PRL81}
\bibinfo{author}{\bibfnamefont{D.}~\bibnamefont{Scalapino}} \bibnamefont{and}
  \bibinfo{author}{\bibfnamefont{R.}~\bibnamefont{Sugar}},
  \bibinfo{journal}{Phys. Rev. Lett.} \textbf{\bibinfo{volume}{46}},
  \bibinfo{pages}{519} (\bibinfo{year}{1981}).

\bibitem[{\citenamefont{Blankenbecler et~al.}(1981)\citenamefont{Blankenbecler,
  Scalapino, and Sugar}}]{BSS81}
\bibinfo{author}{\bibfnamefont{R.}~\bibnamefont{Blankenbecler}},
  \bibinfo{author}{\bibfnamefont{D.~J.} \bibnamefont{Scalapino}},
  \bibnamefont{and} \bibinfo{author}{\bibfnamefont{R.~L.} \bibnamefont{Sugar}},
  \bibinfo{journal}{Phys. Rev. D} \textbf{\bibinfo{volume}{24}},
  \bibinfo{pages}{2278} (\bibinfo{year}{1981}).

\bibitem[{\citenamefont{Rombouts et~al.}(1999)\citenamefont{Rombouts, Heide,
  and Jachowicz}}]{Romb99}
\bibinfo{author}{\bibfnamefont{S.~M.~A.} \bibnamefont{Rombouts}},
  \bibinfo{author}{\bibfnamefont{K.}~\bibnamefont{Heide}}, \bibnamefont{and}
  \bibinfo{author}{\bibfnamefont{N.}~\bibnamefont{Jachowicz}},
  \bibinfo{journal}{Phys. Rev. Lett.} \textbf{\bibinfo{volume}{82}},
  \bibinfo{pages}{4155} (\bibinfo{year}{1999}).

\bibitem[{\citenamefont{Rubtsov}()}]{Rub}
\bibinfo{author}{\bibfnamefont{A.~N.} \bibnamefont{Rubtsov}},
  \eprint{cond-mat/0302228}.

\bibitem[{\citenamefont{DeTar and Gottlieb}(2004)}]{QCD}
\bibinfo{author}{\bibfnamefont{C.}~\bibnamefont{DeTar}} \bibnamefont{and}
  \bibinfo{author}{\bibfnamefont{S.}~\bibnamefont{Gottlieb}},
  \bibinfo{journal}{Physics Today} \textbf{\bibinfo{volume}{57}},
  \bibinfo{pages}{45} (\bibinfo{year}{2004}).

\bibitem[{\citenamefont{Fetter and Walecka}(1971)}]{Fetter-Walecka}
\bibinfo{author}{\bibfnamefont{A.~L.} \bibnamefont{Fetter}} \bibnamefont{and}
  \bibinfo{author}{\bibfnamefont{J.~D.} \bibnamefont{Walecka}},
  \emph{\bibinfo{title}{Quantum theory of many-particle systems}}
  (\bibinfo{publisher}{McGraw-Hill}, \bibinfo{address}{New York},
  \bibinfo{year}{1971}).

\bibitem[{\citenamefont{Prokof'ev and Svistuniov}(1998)}]{DMC}
\bibinfo{author}{\bibfnamefont{N.~V.} \bibnamefont{Prokof'ev}}
  \bibnamefont{and} \bibinfo{author}{\bibfnamefont{B.~V.}
  \bibnamefont{Svistuniov}}, \bibinfo{journal}{Phys. Rev. Lett.}
  \textbf{\bibinfo{volume}{81}}, \bibinfo{pages}{2514} (\bibinfo{year}{1998}).

\bibitem[{\citenamefont{Husslein et~al.}(1997)\citenamefont{Husslein, Fettes,
  and Morgenstern}}]{Husslein97}
\bibinfo{author}{\bibfnamefont{T.}~\bibnamefont{Husslein}},
  \bibinfo{author}{\bibfnamefont{W.}~\bibnamefont{Fettes}}, \bibnamefont{and}
  \bibinfo{author}{\bibfnamefont{I.}~\bibnamefont{Morgenstern}},
  \bibinfo{journal}{Int. J. Mod. Phys.} \textbf{\bibinfo{volume}{C8}},
  \bibinfo{pages}{397} (\bibinfo{year}{1997}), \bibinfo{note}{also
  cond-mat/9705026}.

\bibitem[{\citenamefont{Hirsch}(1983)}]{Hirsch83}
\bibinfo{author}{\bibfnamefont{J.~E.} \bibnamefont{Hirsch}},
  \bibinfo{journal}{Phys. Rev. B} \textbf{\bibinfo{volume}{28}},
  \bibinfo{pages}{4059} (\bibinfo{year}{1983}).

\bibitem[{\citenamefont{Sewer et~al.}(2002)\citenamefont{Sewer, Zotos, and
  Beck}}]{Sewer02}
\bibinfo{author}{\bibfnamefont{A.}~\bibnamefont{Sewer}},
  \bibinfo{author}{\bibfnamefont{X.}~\bibnamefont{Zotos}}, \bibnamefont{and}
  \bibinfo{author}{\bibfnamefont{H.}~\bibnamefont{Beck}},
  \bibinfo{journal}{Phys. Rev. B} \textbf{\bibinfo{volume}{66}},
  \bibinfo{pages}{140504} (\bibinfo{year}{2002}).

\bibitem[{\citenamefont{Charutz et~al.}(1997)\citenamefont{Charutz, Neuhauser,
  and Rom}}]{Rom97}
\bibinfo{author}{\bibfnamefont{D.}~\bibnamefont{Charutz}},
  \bibinfo{author}{\bibfnamefont{D.}~\bibnamefont{Neuhauser}},
  \bibnamefont{and} \bibinfo{author}{\bibfnamefont{N.}~\bibnamefont{Rom}},
  \bibinfo{journal}{Chem. Phys. Lett.} \textbf{\bibinfo{volume}{270}},
  \bibinfo{pages}{382} (\bibinfo{year}{1997}).

\bibitem[{\citenamefont{Baeurle}(2002)}]{Baeurle02}
\bibinfo{author}{\bibfnamefont{S.~A.} \bibnamefont{Baeurle}},
  \bibinfo{journal}{Phys. Rev. Lett.} \textbf{\bibinfo{volume}{89}},
  \bibinfo{pages}{080602} (\bibinfo{year}{2002}).

\end{thebibliography}

\end{document}